\documentclass[12pt,a4paper,fleqn]{article}
\pdfoutput=1

\usepackage{epsfig}
\usepackage{amsmath,amssymb,latexsym,float}
\usepackage{subfigure}
\usepackage{authblk}
\usepackage{anysize}
\usepackage{achicago}
\usepackage{geometry}
\usepackage{color}
\usepackage{float}

\newcommand{\ie}{i.e. }

\newcommand{\etal}{\textit{et al. }}
\newcommand{\es}{\textsc{ESPResSo }}
\newcommand{\esNoSpace}{\textsc{ESPResSo}}

\title{Modelling nano-particle agglomeration using local interactions}
\author[1]{G. Inci}
\author[2]{A. Arnold}
\author[1]{A. Kronenburg\thanks{Corresponding author. Current address: Institute for Combustion Technology, University 
of Stuttgart, Herdweg 51, 70174 Stuttgart, Germany. Telephone: +49 711 685-65635; Fax: +49 711 685-55635 Electronic mail: kronenburg@itv.uni-stuttgart.de}}
\author[2]{R. Weeber}

\affil[1]{Institute for Combustion Technology, University of Stuttgart, Herdweg 51, 70174 Stuttgart, Germany}
\affil[2]{Institute for Computational Physics, University of Stuttgart, Pfaffenwaldring 27, 70569 Stuttgart, Germany}

\begin{document}

\maketitle

\begin{abstract}
Nano-particle agglomeration plays an important role in processes such as spray drying and particle flame synthesis. These processes have in common that nano-particles collide at low concentrations and get irreversibly linked at the point of contact due to plastic deformation. In this paper, we investigate several models of irreversible connections, which require only local interactions between the colliding nano-particles and thus allow for scalable simulations. The models investigated here connect the particles upon collision by non-bonded strongly attractive interactions, bonded interactions or by binding agents placed at the point of contact. Models using spherically symmetric interactions form compact agglomerates and are therefore unsuitable to study agglomeration. In contrast, models that are either based on both central and angular potentials (type one) or on binding agents (type two) efficiently prevent restructuring of the agglomerates, and are therefore useful for modeling contacts formed by plastic deformation. Moreover, both types of models allow to control the rigidity and by that the degree of restructuring. The first type of model is computationally more efficient at low fractional dimensions of the aggregates, while the second gives easy access to local shear forces, which is important when breaking of agglomerates is to be considered. As example applications, we reproduce the well-known diffusion-limited agglomeration (DLA) and report results on soot aggregation.

\end{abstract}


%\pagenumbering{gobble}

%% main text
\section{Introduction}

Particle agglomeration is a process in which small particles with a diameter of 1--100\,nm collide and stick to form larger agglomerates. Agglomeration is the main growth mechanism in spray drying and particle flame synthesis, but also plays an important role in water purification, mineral beneficiation and biological separation processes \shortcite{kodas,gregory}.
Experimental studies provide information on the size and shape of selected agglomerates, but not on the history of the agglomeration process or the intermolecular forces within the agglomerates that are of importance for the dynamics of the formation of the final product \shortcite{forrest,witten,mulholland1988,friedlander}. Computer simulations can help to gain more insight, since they allow for tracking of individual particles and by that provide a better understanding of the agglomerate's growth and morphology.

% modeling questions

The relevance of simulations crucially depends on the underlying model, since the agglomerating particles can be of very different nature, from crystallites to small protein aggregates. The particles feel short-range attraction due to induced dipole forces that results in an elastic deformation and thus strengthens the contact, or they create bridges from one particle to another~\shortcite{marshall2011,gay,kinloch}. Usually, this attraction is so strong that the contacts can be considered permanent.
In addition, the particles are dispersed in a fluid that mediates flow and  Brownian motion~\shortcite{zaichik,garrick,babler2005}, so that particles adhere only when the attractive force overcomes the thermal fluctuations and the hydrodynamic drag \shortcite{sander,vanni2002}. The properties of the formed agglomerates therefore depend both on the modeling of the attractive particle interactions and the fluid.

% known simulation results 

Peng \etal \citeyear{doroodchi} used a discrete element method (DEM) to simulate the nanoparticle agglomeration due to random Brownian diffusion. Even though the simulation results were found to be in good agreement with experimental data, the coordination numbers were too high. Binder \etal \citeyear{peukert} showed that hydrodynamic interactions lead to compact agglomerates that are exposed to high drag forces. Recently, Isella and Drossinos \citeyear{drossinos} investigated numerically the agglomeration of spherically symmetric nanoparticles that undergo Brownian motion and stick irreversibly when coming into contact. The aggregates were found to be compact, tubular and elongated.

% compact clusters are due to sliding

These numerical studies highlight a common issue, namely that the agglomerates obtained from simulations are often too compact. The reason for this is sliding and/or the rolling of particles within an agglomerate. This rearrangement, which leads to rather compact agglomerates as time proceeds, may or may not be physical. Particles that are tied together by weak physical van der Waals forces (soft agglomerates) may restructure when heated or exposed to high tension \shortcite{schmidt}. However, the particle bonds within hard agglomerates are sufficiently strong and restructuring after collision does not appear, so that compact clusters are not found.
This may also be true for soft agglomerates such as agglomerates of real silica or titania that tend to break rather than to restructure \shortcite{dominik1997}.

% need to prevent sliding

Thus in order to model strongly binding particles in computer simulation, contact bonds that prevent rearrangement are necessary. Kusaka \etal \citeyear{kusaka}, Kempf \etal \citeyear{kempf}, and Iglberger \etal \citeyear{iglberger} coupled the nanoparticle motion models with a rigid body motion of the formed aggregates. Although rigid body dynamics solves the undesired restructuring, it is a non-local interaction that requires to exchange data between all processors that a particular agglomerate occupies, and thus limits the scalability to large systems.

In our study, we investigate several models of particle binding that rely only on local interactions, and by that naturally allow for scalable simulations. The contact binding is achieved by simple van der Waals attraction, formation of harmonic and/or angular bonds, or by binding agents at the contact points, so that the particles loose their rotational symmetry (compare Table~\ref{tab:algorithm}). The Brownian motion of the particles is modeled by Langevin dynamics~\shortcite{Limbach2006704}, and for simplicity, the particles only feel a weak van der Waals attraction, apart from the contact binding interactions. We show that central potentials always lead to compact agglomerates, independently of whether they are introduced as an always present non-bonded interaction with finite well depth or as harmonic bonds. In contrast, models that either introduce angular bonds or binding agents and thus break rotational symmetry can very effectively prevent restructuring at low computational cost. Moreover, the models allow to control the rigidity of the agglomerates and thus control restructuring accurately.

This paper is organized as follows: Section~\ref{sec:theory} introduces the potentials between the particles and the binding models. Section~\ref{sec:results} reports simulation results on the fractal and morphological properties of the agglomerates for each model. Section~\ref{sec:results} then investigates the non-central interaction models in more detail, and compares present results to known results for the classical diffusion limited aggregation (DLA) and to results of other numerical and experimental studies on the formation and growth of soot aggregates.

\section{Numerical Methods}
\label{sec:theory}

Simulations are performed using the Molecular Dynamics (MD) software package \es \shortcite{espresso2012}. It provides all the required interactions plus a number of additional features such as electrostatic or hydrodynamic interactions, that might be of interest for future studies of agglomeration. 

In this section, we first describe the interactions between the particles, then explain our models of binding of the particles that have collided and finally give details on our test simulations such as the equations of motion employed.

\subsection{Relevant interactions between particles}
\label{subsec:interactions}

The potential between two primary particles is modelled by integrating the intermolecular potential over the volume of these particles. The traditional (12-6) Lennard-Jones potential is used to model the intermolecular potential. It combines a long-range attractive force (the $1/r^6$-term) with a short-range repulsive force (the $1/r^{12}$-term) between the particles,

\begin{equation}
\label{eq:LJ}
  V_{LJ}(r)=4\epsilon \left ( \left ( \frac{\sigma_{LJ} }{r} \right )^{12} - \left ( \frac{\sigma_{LJ} }{r} \right )^{6}\right )
\end{equation}

where $r$ is the particles' distance, $\sigma_{LJ}$ is the distance at which the potential equals zero and $\epsilon$ is the depth of the attractive potential, defining the maximum attractive energy between two particles. 

The potential between two particles ($U(r)$) is obtained from the integration of Eq.~\ref{eq:LJ} over two spherical particles of diameter $\sigma$ \shortcite{potential-1,potential-2}.  It is given by 

\begin{subequations}\label{eq:integrated-LJ}
\begin{align}
U_{LJ}^{attractive}(r)&=&-\frac{A}{6}\left( ln \left( \frac{r^2-\sigma ^2}{r^2} \right) + \frac{\sigma ^2}{2(r^2-\sigma ^2)} + \frac{\sigma ^2}{2r^2} \right), \label{eq:LJ-attractive}\\
 U_{LJ}^{repulsive}(r)&=&\frac{A\sigma _{LJ}^6}{2520r} \lbrace \sigma ^2 \left( \frac{1}{2(r-\sigma)^7} + \frac{1}{2(r+\sigma)^7} + \frac{1}{r^7} \right) \nonumber \\ 
 && -\frac{\sigma}{3} \left( \frac{1}{(r-\sigma)^6} - \frac{1}{(r+\sigma)^6} \right)\nonumber \\
 &&- \frac{1}{15} \left( \frac{2}{r^5} - \frac{1}{(r-\sigma)^5} -\frac{1}{(r+\sigma)^5} \right) \rbrace , \label{eq:LJ-repulsive}\\
   U(r)&=&U_{LJ}^{attractive}(r)+U_{LJ}^{repulsive}(r).
\end{align}
\end{subequations}
In Eq.~\eqref{eq:integrated-LJ}, A is the Hamaker constant and expressed by
\begin{equation}
\label{hamakar-constant}
A=4\pi\epsilon\sigma_{LJ}^6n^2,
\end{equation} 
where $n$ is the molecular number density in the solid \shortcite{drossinos}.

In the present study, the interaction between two clusters is obtained via the particle potentials between the respective particles that belong to the different clusters and are closer in distance than a cut-off distance $r_{cut}$. In addition, adjacent particles can be bound via a 2-body spring potential and 3-body angular potentials. The 2-body spring bond (harmonic bond) potential between two particles in contact is given by
\begin{equation}\label{eq:2body-harmonic}
U_{harmonic-bond} = \frac{1}{2} k_h (r_{ij} - r_0)^2 ,
\end{equation}
where $ r_{ij} = \Vert\vec{r}_j - \vec{r}_i\Vert$ gives the distance between the particles, $ r_0$ is the equilibrium distance, and $ k$ is the spring constant. The 3-body angular bond is used to fix the angle between the position vectors from the center particle to the two other particles. The energy configuration of the angular bond potential is
\begin{equation}\label{eq:angle}
U_{angle} = k_a (\theta - \theta_0)^2 , 
\end{equation}
where $ \theta$ is the angle in radians between vectors $ \vec{r}_{ij} = \vec{r}_j - \vec{r}_i$ and $ \vec{r}_{kj} = \vec{r}_j - \vec{r}_k$, $ \theta_0$ is the equilibrium angle, and $k$ is the angle bond constant. 

\subsection{Equations of motion}

Langevin dynamics (LD) is used to model the particle motion including Brownian motion. We use a Velocity-Verlet-integrator plus a Langevin thermostat to integrate the Langevin equation \shortcite{Limbach2006704}. At each time step all particles are subjected to a random force and a frictional force such that these two forces satisfy the fluctuation-dissipation theorem and balance each other. In this formalism, the Langevin equation for the i-th particle \shortcite{rkubo} is given by
\begin{equation}
\label{eq:Langevin}
  m\dot{\textbf{u}}_{i}(t)={\textbf{F}_C}_i -\gamma {\textbf{u}}_{i} + \textbf{W}_{i}(t) ,
\end{equation}
where ${\textbf{F}_C}_i$ are conservative forces that arise from inter-particle potentials that the $i$-th particle feels (${\textbf{F}_C}_i$=-\textbf{$\bigtriangledown$} $U_i$), $\gamma$ is the friction constant between the primary particles and surrounding fluid, $u_{i}$ is the velocity of the i-th particle and $W_{i}(t)$ is Einstein's white noise term, which is a Gaussian random source \shortcite{fox,lemons}. The noise models the random kicks of the fluid molecules to the particles with zero-mean and satisfies
\begin{eqnarray}\label{eq:noise}
  <W_{i}(t)>=0,&
  &\left\langle W(t)W(t') \right\rangle = 2\gamma k_B T\delta(t-t').
\end{eqnarray}

 The exact modelling of the friction coefficient may also depend on the Knudsen number $Kn$, which is the ratio of the fluid molecules mean free path length $l_{fluid}$ to the particle radius, $Kn= 2 l_{fluid}/\sigma$ \shortcite{knudsen-weber}. For large particles ($Kn<<1$), the flow is in the continuum regime and the well known Stokes law applies, viz.
\begin{equation}\label{eq:f_second}
  \gamma=3\pi\mu_{f}\sigma,
\end{equation}

with $\mu_f$ denoting the dynamic viscosity of the surrounding fluid. For smaller particles ($Kn>1$), however, a slip velocity between the particle and the surrounding fluid will exist. The friction coefficient will be lower than the Stokes law would predict and the so-called Cunningham correction, $C_{C}$, needs to be introduced as is indicated by Eqs. ~\eqref{eq:f_third} and ~\eqref{eq:f_fourth} \shortcite{chen},
\begin{subequations}\label{eq:friction}
\begin{align}
  \gamma =\frac{3\pi \mu_{f} \sigma}{C_{C}}\label{eq:f_third}\\
  \text{with }C_{C}=1+Kn\left ( A_{1}+A_{2} \cdot  \mathrm{exp} \left [ \frac{-2A_{3}}{Kn} \right ] \right )\label{eq:f_fourth},
\end{align}
\end{subequations}
where the constants $A_1, A_2$ and $A_3$ are 1.257, 0.4 and 0.55, respectively \shortcite{chen}.

Some of the models that are described in the next section (sec. \ref{models}) require rotationally non-invariant particles. Their orientations can be expressed by quaternions \shortcite{martys} that are defined in terms of Euler angles for an individual particle as
\begin{eqnarray}\label{eq:quaternion-1}
q_0&=&cos(\beta/2)cos((\alpha+\psi)/2),\nonumber\\
q_1&=&sin(\beta/2)cos((\alpha-\psi)/2),\nonumber\\
q_2&=&sin(\beta/2)sin((\alpha-\psi)/2),\nonumber\\
q_3&=&cos(\beta/2)sin((\alpha+\psi)/2)
\end{eqnarray}
and satisfy the equality
\begin{equation}\label{eq:quaternion-2}
  q_0^2+q_1^2+q_2^2+q_3^2=1.
\end{equation}
The time derivative of the quaternions can then be expressed via the principal angular velocity $\textbf{w}_p$ as
\begin{equation}\label{eq:quaternion-3}
\dot{\textbf{q}}=\begin{pmatrix}
\dot{q_0}\\ 
\dot{q_1}\\ 
\dot{q_2}\\ 
\dot{q_3}
\end{pmatrix} =\begin{pmatrix}
 -q_2 & -q_3 & q_1 & q_0 \\ 
 q_3 & -q_2 & -q_0 & q_1 \\ 
 q_0 & q_1 & q_3 & q_2 & \\ 
 -q_1 & q_0 & -q_2 & q_3 
\end{pmatrix} \begin{pmatrix}
\textbf{w}_{px} \\ 
\textbf{w}_{py} \\ 
\textbf{w}_{pz} \\ 
0
\end{pmatrix}.
\end{equation} 
This equation for the quaternions is integrated using the same velocity--Verlet integrator as for the solution of the momentum equation.  The necessary thermalization of the rotational degrees of freedom is based on Langevin dynamics and is equivalent to the procedure used for the translational degrees of freedom. More details can be found in \shortcite{martys}.

\subsection{Particle agglomeration models}
\label{models}

Two particles are assumed to link permanently and form a cluster when their center-to-center distance, $r$, becomes smaller than the particle diameter $\sigma$. Clusters with at least one common particle constitute an agglomerate. \es does not explicitly identify agglomerates. We therefore use an external postprocessing tool that performs cluster labeling based on bond information from \esNoSpace. Particles are processed serially. If a particle is not connected to a (preceding) particle that had been identified as part of a cluster, this particle forms a new cluster. Otherwise the particle inherits the cluster label of the particle it is connected to. If the particle is connected to more than one particle with different cluster labels, these clusters are merged by assigning a single cluster label to all particles within the clusters.
%and an additional bond list needs to be compiled from the standard bond information given in \es by a linked TCL script that could be interpreted as a postprocessing tool. The list is updated at selected time steps for an unambiguous assignment of the agglomerates to their respective primary particles.} 

In this paper, four different models that simulate the simultaneous formation and growth of agglomerates, are compared. For the first two models, only the adhesion force between the contacting particles is considered. However, this force acts along the line connecting the centers, and it therefore does not exert torque on the spherical particles. In a further step, we therefore introduce a range of forces and torques, in order to avoid particles from sliding and/or rolling motions. These models are summarized in Table~\ref{tab:algorithm} and described in the 4 following subsections.

\begin{table*}[t]
\begin{center}
\caption{Summary of the four different agglomeration formation models examined in this report.}
    \begin{tabular}{ | p{1.5cm} | p{2.8cm} | p{3.8cm} | l |}
    \hline
    Model Number & Model Name & Model Description & Acronym \\ \hline
    i & Non-bonded potential model & collision model only with non-bonded potential & NB \\ \hline
    ii & Single bond model & single bond potential between collided particles model & SB \\ \hline
    iii & All bonds model & single bond and angular bond potentials model & AB \\ \hline
    iv & Angle bonded virtual sites model & collision model with two virtual particles connected through angle potential & AnBV \\ \hline
    \end{tabular}
\label{tab:algorithm}
\end{center}
\end{table*}

\subsubsection{Model using non-bonded potentials (NB)}
\label{sec:LJ-model}
In this model, when particles come close, a potential (Eq.~\eqref{eq:integrated-LJ}) with a highly attractive well (such as $\tilde{A}=190$ as given in Table~\ref{tab:parameters} below) would make them very unlikely to separate. However, the interactions are still rotationally invariant. 

\subsubsection{Single bond model (SB)}

For the "single bond" model, a distance-dependent bond potential (\ie harmonic bond potential) is used to ensure a binding energy between two contacting particles according to the JKR theory \shortcite{chokshi}. The binding energy due to the surface deformation is given by
\begin{equation}
\label{eq:jkr}
E_{adh}^{JKR}=\Psi\pi a^2,
\end{equation}
where $a$ is the radius of the circular contact region and expressed by 
\begin{eqnarray}
\label{eq:jkr-2}
a=\left( \frac{3\Psi\pi (\sigma/2)^2}{2K}\right) ^{1/3},\nonumber \\ K=\frac{2E_Y}{3(1-\upsilon ^2)}, \nonumber \\
\Psi\approx\frac{A}{24\pi D_0^2}.
\end{eqnarray}
Here, $\Psi$ is the surface energy of the particles, $\upsilon$ is Poisson's ratio, $E_Y$ is Young's modulus and $D_0 = 0.165$ nm \shortcite{furst}. In Eq.~\eqref{eq:jkr-2}, the Hamaker constant $A$ is estimated from experimentally measured fragmentation curves for diesel soot \shortcite{kasper,chokshi}. Eqs.~\eqref{eq:jkr}--\eqref{eq:jkr-2} can thus be used to relate the modelled bond potential in Eq.~\eqref{eq:2body-harmonic} with the actual physical process by $k_h=2E_{adh}^{JKR}/a^2$. The equilibrium distance is set to 
$r_0=\sigma$ which ensures direct particle contact.

\subsubsection{All bonds model (AB)}

The "all bonds" model uses 3-body angular bond potentials to represent tangential forces in addition to the 2-body spring force. In this model, the 2-body potential is used to keep particles together, and the 3-body potentials are responsible for avoiding the sliding/rolling of connected particles. Similar to Becker and Briesen's study \citeyear{becker} the angle bond constant $k_a$ (cf. Eq.~\eqref{eq:angle}) is estimated using a single-bond bending rigidity, viz.
\begin{equation}
\label{eq:bond-constant-k}
k_a=\frac{\kappa_0}{192},
\end{equation}
where $\kappa_0$ is the bending rigidity per bond and given by \shortcite{furst}
\begin{equation}
\label{eq:bending-rigidity}
\kappa_0=\frac{3\pi a^4E_Y}{4(\sigma/2)^3}.
\end{equation}

\subsubsection{Angle bonded virtual particles model (AnBV)}

Our goal is that the particles should always be connected at the particular point where they initially touched. This can be achieved by using so-called virtual particles, \ie the particles' evolution is not subject to the integrated equations of motion, but they are placed at the surface of a given real particle and its relative position to this particle is fixed. This requires that particles carry a co-moving reference frame that is integrated by rotational Langevin dynamics. In our model, the virtual particles are dynamically created during the simulation and move with the Brownian particles they are attached to. 
%\chdbl{with the equation given by
%\begin{equation}
%\label{eq:virtualsites}
%\textbf{x}_v=\textbf{x}_p+O_p(O_v+\textbf{E}_z)d,
%\end{equation}
%where $\textbf{x}_n$ is the position of the real particle, $O_n$ is the orientation of the real particle, $O_v$ indicates the orientation of the vector $\textbf{x}_v-\textbf{x}_p$ with respect to the real particle's body fixed frame and $d$ is the distance between virtual and real particle}.

\begin{figure}
\center{\includegraphics[scale=0.155]{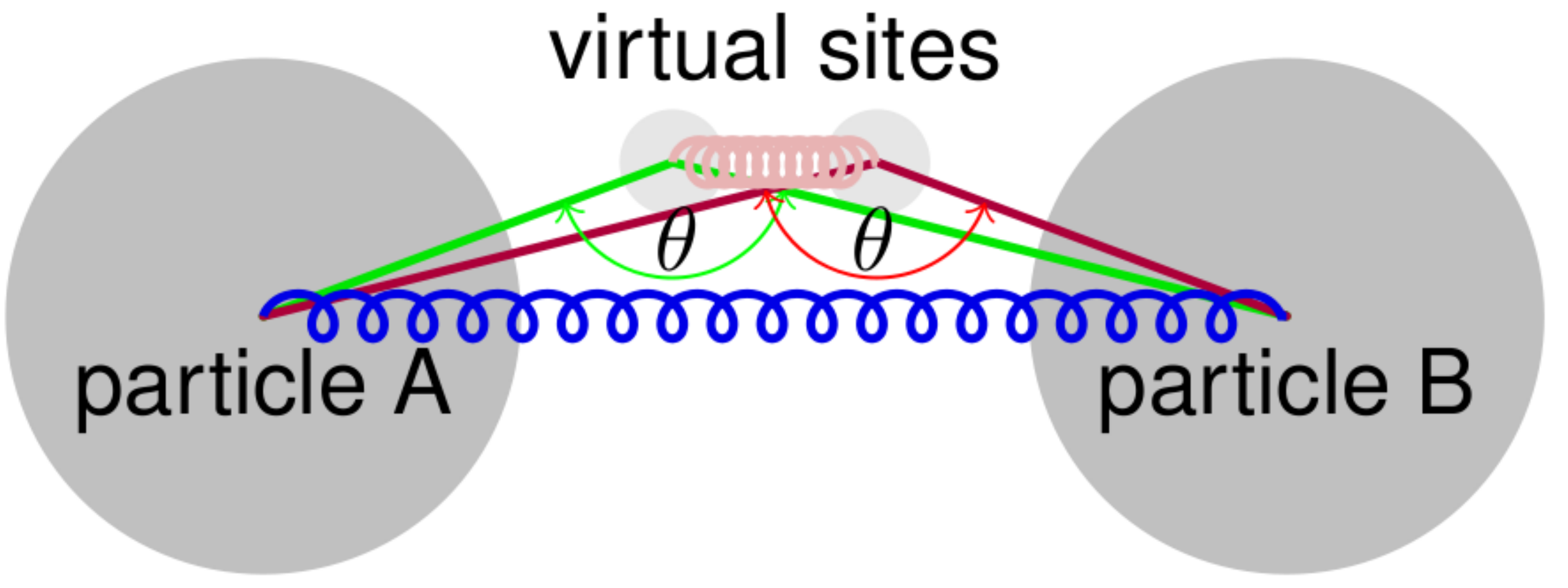}}
\caption{AnBV model, two virtual particles are placed at the collision point and the angle potential (with $\theta_0=\pi$) is defined by two collided particles and virtual particles. Note that we use an exploded view for better illustration.}
\label{fig:anglevirtual}
\end{figure}

 For the "angle bonded virtual particles" model, two virtual particles are created at the point of contact and fixed in the reference frame of the two colliding particles. They are connected to both particles that have collided by angle-dependent potentials with angle $\theta_0 = \pi$ (see equation~\eqref{eq:angle}) as illustrated in Figure~\ref{fig:anglevirtual}.

A 2-body spring bond potential (cf. equation~\eqref{eq:2body-harmonic}) with bond length $\sigma$ between the centers of the particles that collided and a second spring with zero bond length between the virtual sites prevent significant motion around the point of contact, and the angular potential through particles and virtual particles prevents the particles from sliding around each other. This model now is part of the opensource distribution of \es as a general dynamic bonding feature. Documentation on the creation of virtual particles and dynamic bonding can be found in the \es user's guide. 

\subsection{Simulation parameters}
\label{sec:sim-parameters}

In this study, the simulations are carried out with a system consisting of spherical particles placed in a cubic box with periodic boundary conditions for all sides. The surrounding fluid is assumed to be air. 

 We use "reduced units", in which three reference quantities define the unit system. These are the units of energy ($\epsilon^*$), length ($\sigma^*$) and mass ($m^*$). All other quantities are expressed in units that can be derived from $\epsilon^*$, $\sigma^*$ and $m^*$. In the following, the reference values of energy, length and mass are set to the thermal energy of the system, the diameter of the particle, and the mass of the particle, respectively. The normalized key quantities of the LD simulation are then calculated by dividing the real values by the reference units (e.g. $\widetilde{\tau}=\tau/(\frac{\sigma^{*2}m^*}{\epsilon^*})^{\frac{1}{2}}$, $\widetilde{A}=A/\epsilon^*$). 

 The time step ($\widetilde{\tau}=\tau/\tau^*$) of the LD simulations plays an important role in solving the equation of particle motion. For the reduction of the computational time it should be as large as possible and preserve accuracy. To obtain reliable simulations, the simulation time step should be much smaller than the relaxation time of the particles, which is the characteristic time for a particle to adjust its velocity. The relaxation time of a Brownian particle is given by

\begin{equation}
\label{eq:relaxationtime}
\tau_{B}=\frac{m}{\gamma } = \frac{\rho d_{p}^{2} C_{C}}{18 \mu_f }.
\end{equation}

The normalized time step of all test case simulations is set to $\widetilde{\tau}$=0.01 which is smaller than the relaxation time of the Brownian particles, and air at $T=600K$ is typically assumed as the surrounding fluid.

The cut-off distance of $r_{cut}=5\sigma$ has been introduced and beyond this distance the attraction between particles is negligible compared to the thermal energy. The non-dimensionalized value of the Hamaker constant ($\tilde{A}=19$) corresponds to a dimensional quantity of $A=2.38*10^{-19}J$  which is a typical value derived for soot particles \shortcite{kasper}. The corresponding values for the harmonic and the angular bond stiffness constants have been computed from Eqs.~(\ref{eq:jkr-2})--(\ref{eq:bending-rigidity}), assuming the Young's modulus and Poisson's ratio of a soot particle to be $E_Y=2$ GPa and $\nu=0.3$, respectively \shortcite{biswass}. The binding energy due to surface deformation of the two contacting particles with a diameter of $\sigma=50$ nm is calculated to approximate $E_{JKR}=1.3*10^{-17}$ J. This energy is used to estimate the harmonic and angular bond stiffness constants $k_h$ and $k_a$. This leads to a harmonic bond constant for the SB model of $k_h=0.7$ $J/m^2$ assuming $r_{ij}=\sigma-a$ and $r_0=\sigma$ (compare Eqs.~\eqref{eq:2body-harmonic}--\eqref{eq:jkr-2}). The corresponding values for the the AB and AnBV models can be estimated to be $k_h=0.35$ $J/m^2$ and $k_a=0.002$ $J/m^2$, respectively. This assumes an even distribution of the binding energy to the angular and harmonic bonds. The bond stiffness constants are then normalized by the reference quantities ($t^{*2}/m^*$). Note that the derived bond constants are rather large and could be reduced by one order of magnitude. This does not affect the simulation results due to the irreversible nature of the bonds but removes unnecessary numerical stiffness.

The other simulation parameters for the different cases are listed in Table~\ref{tab:parameters}. For all simulations, the canonical ensemble is assumed, where the number of particles N, the volume V and the temperature T are fixed.

\begin{table*}[h!]
\begin{center}
\caption{Simulation parameters}
    \begin{tabular}{ | p{1.8cm} | p{2.1cm} | p{2.1cm} | p{2.1cm} | p{2.1cm} |}     
	\hline
     & Cross-Shape &  Randomly Placed Particles &  DLA & Soot Particle Simulations \\ \hline
    Case \newline Names & Case-1 & Case-2 & Case-3 & Case-4  \\ \hline
    Number of \newline Particles & 17 & 500 &  215 & 1000 \\ \hline
    Normalized Box Length & $36 {\sigma}$ & $40 {\sigma}$ &  $20 {\sigma}$ & $100 {\sigma}$ \\ \hline
    NB Model & $\widetilde{A}$=190 \newline $\sigma=100 nm$ &  \center{-} & \center{-} & - \\ \hline
    SB Model & $\widetilde{A}$=19 \newline $\sigma=100 nm$ \newline $\tilde{k}_h$=1000 & \center{-} &  \center{-} & -\\ \hline
    AB Model & $\widetilde{A}$=19 \newline $\sigma=100 nm$ \newline $\tilde{k}_h$=1000 $\tilde{k}_a$=1000 & \center{-} & \center{-} & - \\ \hline
    AnBV Model & $\widetilde{A}$=19 \newline $\sigma=100 nm$ \newline $\tilde{k}_h$=1000 $\tilde{k}_a$=1000 & $\widetilde{A}$=19 \newline $\sigma=50 nm$ \newline 1$\leq$$\tilde{k}_h$$\leq$1000 10$\leq$$\tilde{k}_a$$\leq$1000  & $\widetilde{A}$=19 \newline $\sigma=100 nm$ $\tilde{k}_h$=500 $\tilde{k}_a$=500  & $\widetilde{A}$=19 \newline $\sigma=20 nm$ $\tilde{k}_h$=500 $\tilde{k}_a$=500 \\ \hline
    \end{tabular}
\label{tab:parameters}
\end{center}
\end{table*}

\section{Results and Discussion}
\label{sec:results}

In this section we present the results of the different test cases summarized in Table ~\ref{tab:parameters}. The first test cases, Case-1 and Case-2, are designed to analyse the capabilities of the different agglomeration models to represent a realistic formation and growth process of agglomerates in dynamic simulations. Case-3 and Case-4 are designed to investigate the contact point model in more detail. 

\subsection{Simulations of the pre-defined agglomerate - Case-1}\label{subsec:Case1}

In this case, the initial positions of particles are given by the shape of a cross. The distance between neighboring particles is closer than the predefined collision criteria in our simulations. Hence, the agglomerates are formed before the iterations start, and the initial shape should be preserved.  

\begin{figure}
\centering
\subfigure[][]{
\includegraphics[scale =0.18] {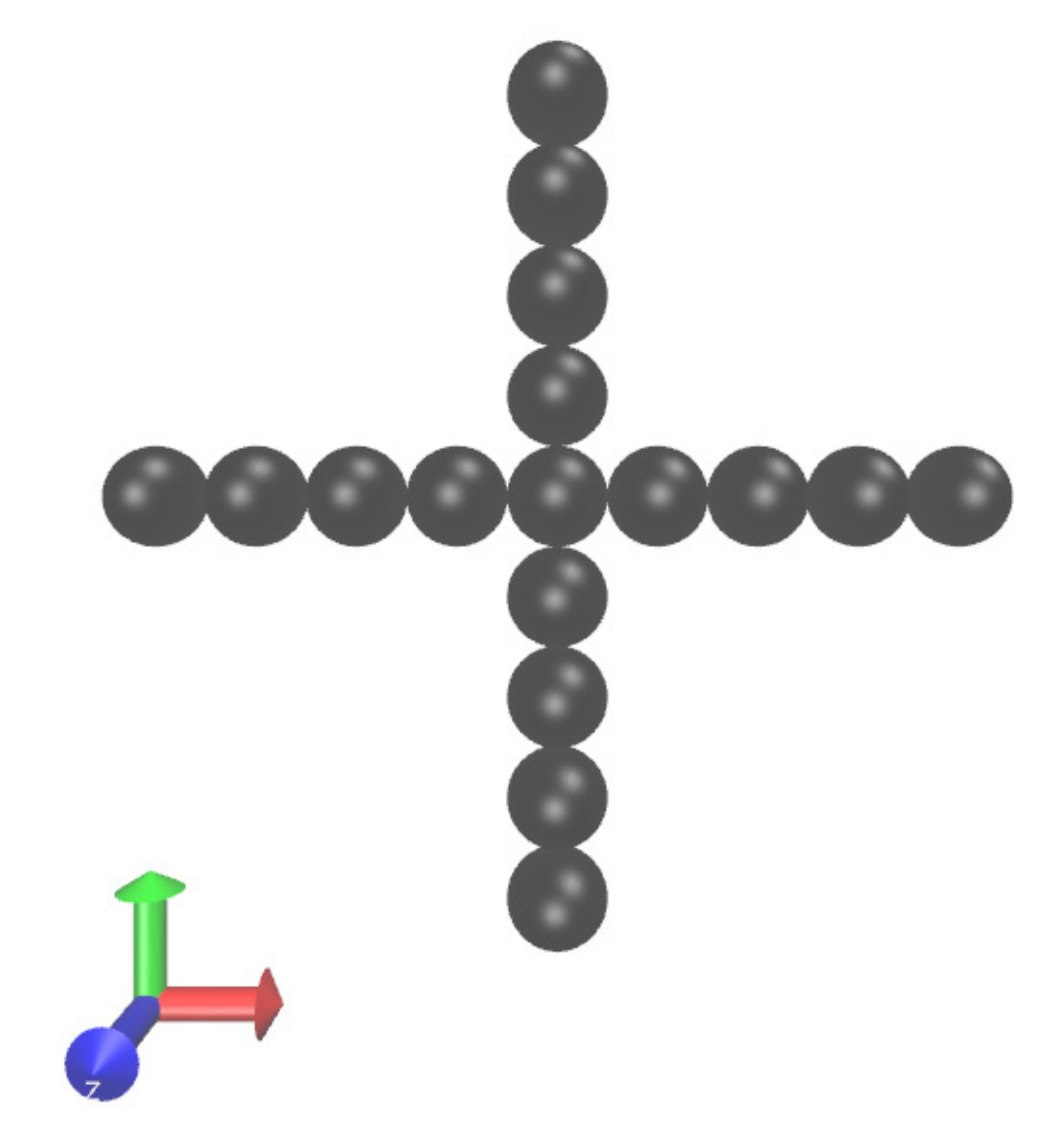}
\label{fig:initial-cross}
}
\subfigure[][]{
\includegraphics[scale =0.18] {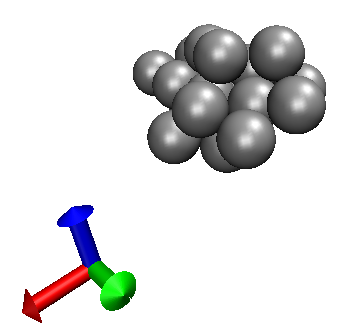}
\label{fig:model-i-1-cross}
}
\subfigure[][]{
\includegraphics[scale =0.18] {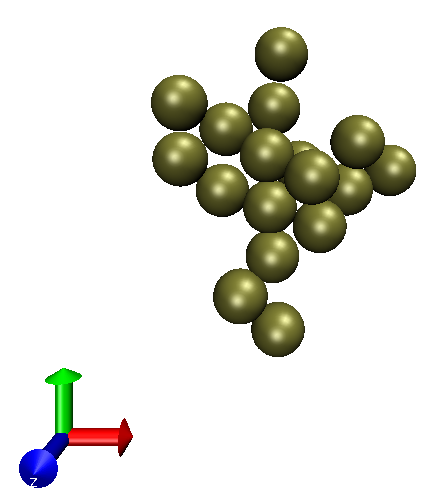}
\label{fig:model-ii-cross}
} \\
\subfigure[][]{
\includegraphics[scale =0.53] {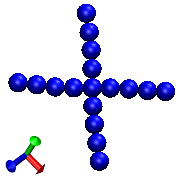}
\label{fig:model-iii-cross}
}
\subfigure[][]{
\includegraphics[scale =0.18] {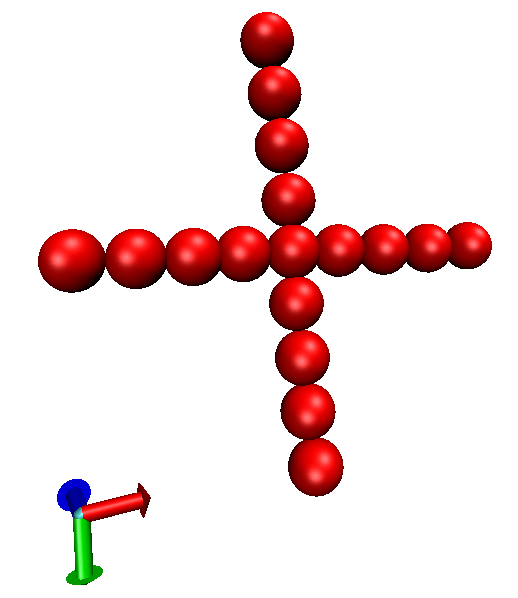}
\label{fig:model-iv-cross}
}
\caption{
\subref{fig:initial-cross} The initial configuration of the cross-shaped agglomerate, and its final configurations obtained from
\subref{fig:model-i-1-cross} NB model with $\widetilde{A}$=190,
\subref{fig:model-ii-cross} SB model,
\subref{fig:model-iii-cross} AB model,
\subref{fig:model-iv-cross} AnBV model simulations. 
}
\label{fig:cross-shape}
\end{figure}

Figure~\ref{fig:cross-shape} shows the configuration of the cross-shaped agglomerate at the end of the simulations using the different models. With the exception of the AB and the AnBV models, the agglomerates do not preserve their initial configuration and change their connectivity. For the NB model, small values of the potential (cf. Eq.~\ref{eq:integrated-LJ}) cannot avoid breakage. Larger Hamaker constants yield rather compact agglomerates ($D_f=3$) that do not preserve the initial shape. 

\begin{figure}
\center{\includegraphics[width=.8\textwidth]{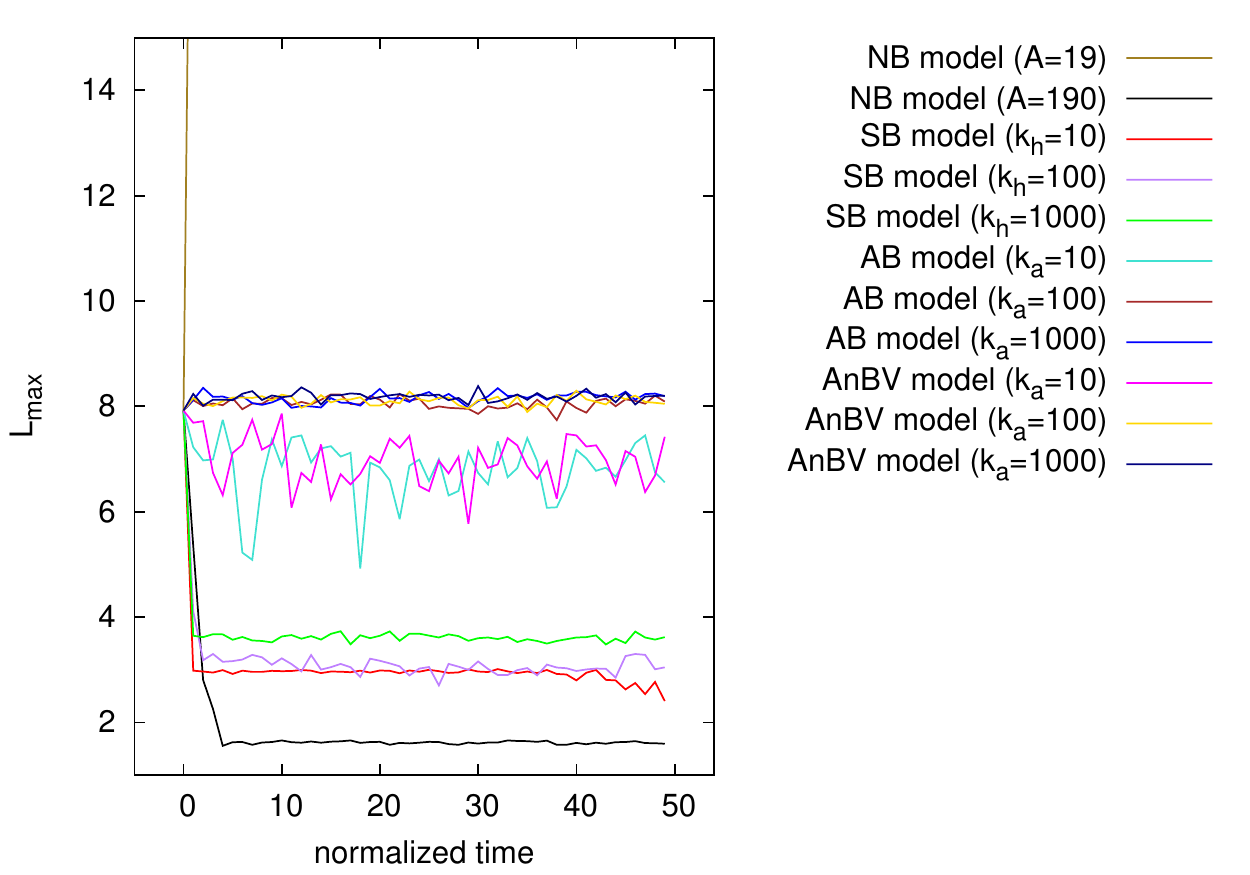}}
\caption{Change in longest distance $L_{max}$ of initially cross-shaped agglomerates for different simulation models and parameters and averaged over 20 simulations. If the cross shape is preserved, $L_{max}$ is constant. Thus, the NB and SB models are unable to preserve the shape independently of parameters. In the AB and AnBV model simulations harmonic bond constant is set as $\tilde{k}_h=1000$, and they require angular bond constant $\tilde{k}_a$ $>$ 10 to preserve the shape.}
\label{fig:cross_ld}
\end{figure}

We use the longest distance ($L_{max}$) between two particles as characteristic quantity to describe shape changes. Figure~\ref{fig:cross_ld} shows its variation with time for all the models used at different parameters. Independent of the Hamakar constant ($A$) and bond strength, the NB and SB models cannot preserve the initial configuration of the clusters, which leads to a very noticeable change in $L_{max}$. The NB model with a low energy minimum leads to the break-up of the agglomerate and $L_{max}$ is determined by the size of the simulation box. Larger potential values, however, result in compact aggregates, so that $L_{max}$ decreases to the distance of around $L_{max}=1.6\sigma$. In contrast, the change in $L_{max}$ is small for the AB and AnBV models with angular bond constants $\tilde{k}_a>10$, which indicates that the shape is preserved. Even for $\tilde{k}_a=10$, $L_{max}$ decreases only slightly, which is in fact only due to increased flexibility of the side arms, while the overall cross shape is still preserved. However, if the agglomerate would be more compact, this flexibility would allow for more contacts and thus more bonds to form. Bond constants can therefore be used to tune the compactification probability.

\begin{figure}
\center{\includegraphics[width=.8\textwidth]{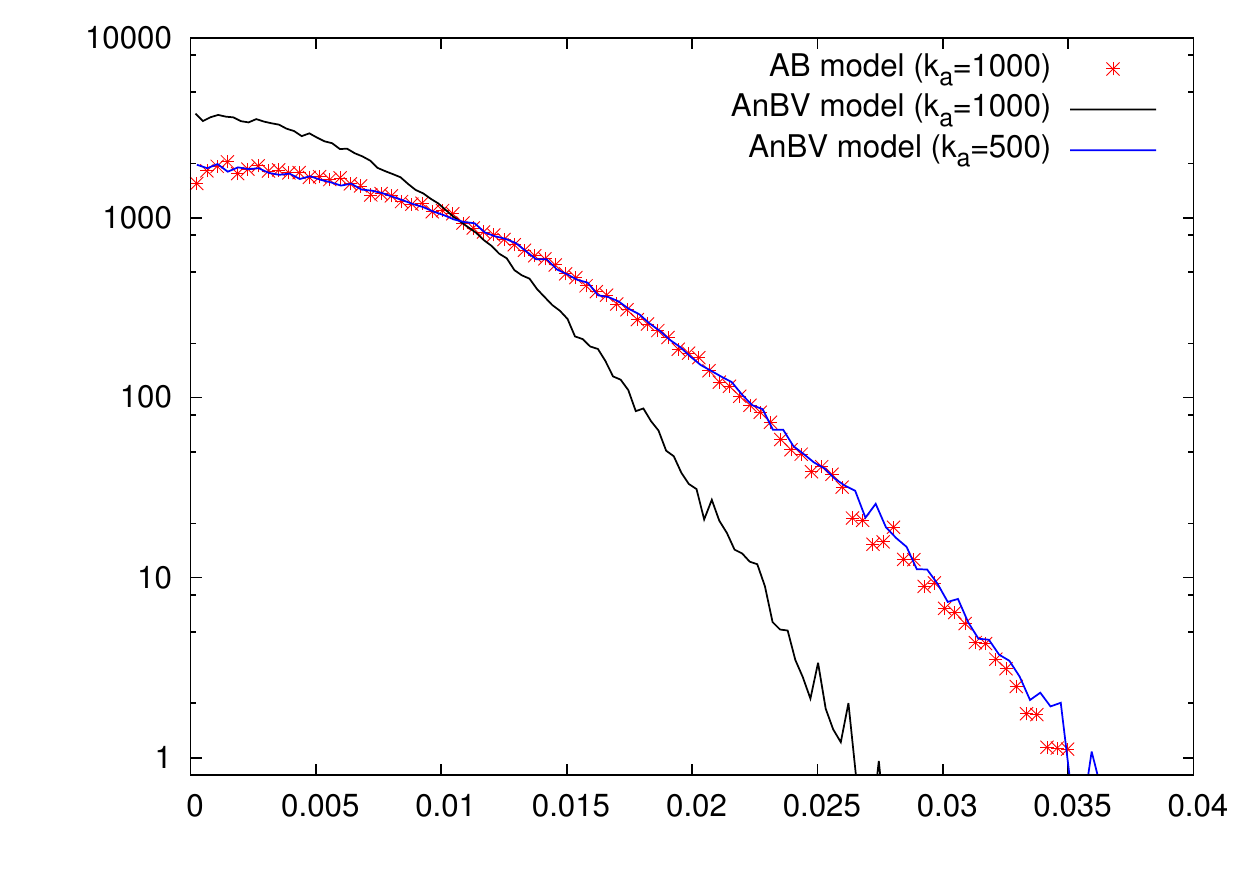}}
\caption{Probability density of the deviation ($\mid \pi-\theta\mid$) from the equilibrium angle between three beads bonded by the AB and the AnBV models. Note that the AB and AnBV models show the same bending characteristics for $\tilde{k}_a=1000$ and $\tilde{k}_a=500$, respectively. This reflects that the AnBV model is constructed using two parallel bonds and therefore requiring only half of the bending rigidity per bond.}
\label{fig:bend}
\end{figure}

Figure~\ref{fig:bend} illustrates the close agreement of the AB and AnBV models. Both models use identical attraction and harmonic bonds to form the bonds. However, for each connection, only one harmonic bond is formed in the AB model, while the AnBV model uses two bonds. Thus, the AnBV model requires just half the bending constant. It is noted here that the dynamic behaviour of the two models can differ slightly (as has been observed for the case of randomly placed particles which is introduced in the next section), since the AnBV model gradually introduces additional degrees of freedom, namely the rotational ones, which are thermalized.  As a consequence the effective temperature of the AnBV model is slightly higher, but the models can still be matched, if the system temperature is reduced accordingly. While the computational results agree well, the models and their implementations are technically distinctly different. The additional virtual particles and the associated bonds lead to an increase in computational time by a factor of two to three for the computation of the resultant forces on the (real) particles and the respective solution of the Langevin equations. However, the angular bond potential describes interactions between three particles in the AB model, while the interactions in the AnBV model are restricted to two physical particles and their respective virtual particles only. Thus, the contact forces between the particles can be directly accessed, and this will be important when large agglomerates (such as soot agglomerats) are subject to shear and breakage is to be considered.

\subsection{Simulations of randomly placed particles - Case-2}\label{subsec:Case2}

Further studies have been performed to investigate the AnBV model's characteristics in more detail and to identify the differences between the AB and AnBV models. The fractal dimension, $D_f$, is now used as a measure of the agglomerates' structure. It is measured from the slope of the double logarithmic plot of the radius of gyration, $R_g$, versus the number of particles, N \shortcite{friedlander2001}. This definition is consistent with the common interpretation of a dimension if the object is compact, i.e. the fractal dimension of a sphere is $D_f$=3, of a disk is $D_f$=2 and of a straight chain is $D_f=1$. Structures that are not compact can be characterised by fractal dimensions.  

The simulations of Case-2 are used for a quantitative and qualitative assessment of the structure of agglomerates that are formed from primary particles subject to Brownian motion. The agglomerates are formed by the collisions of single Brownian particles and/or collisions of clusters during the simulation. Since the clusters are formed by bonded particles, they move naturally following the same unshielded Langevin equation of motion as the nano-particles \shortcite{Isella2011505}. 500 particles have been placed randomly in a box with periodic boundary conditions. These values represent reference values for our sensitivity studies with respect to the bond constants in Sections \ref{subsec:Case1} and \ref{subsec:Case2}.

\begin{figure}
\centering
\subfigure[][]{
\includegraphics[width=.8\textwidth] {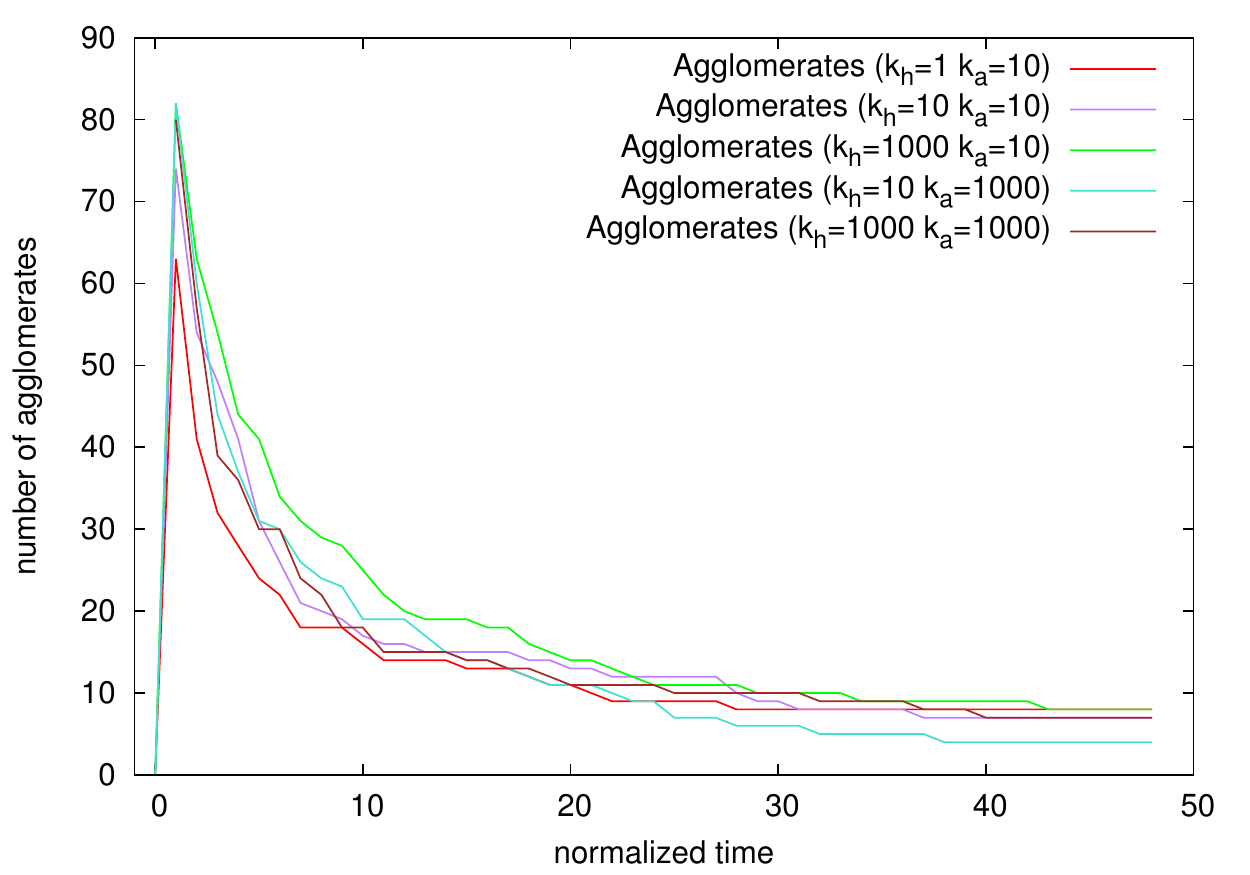}
\label{fig:agg-dynamics}
}
\subfigure[][]{
\includegraphics[width=.8\textwidth] {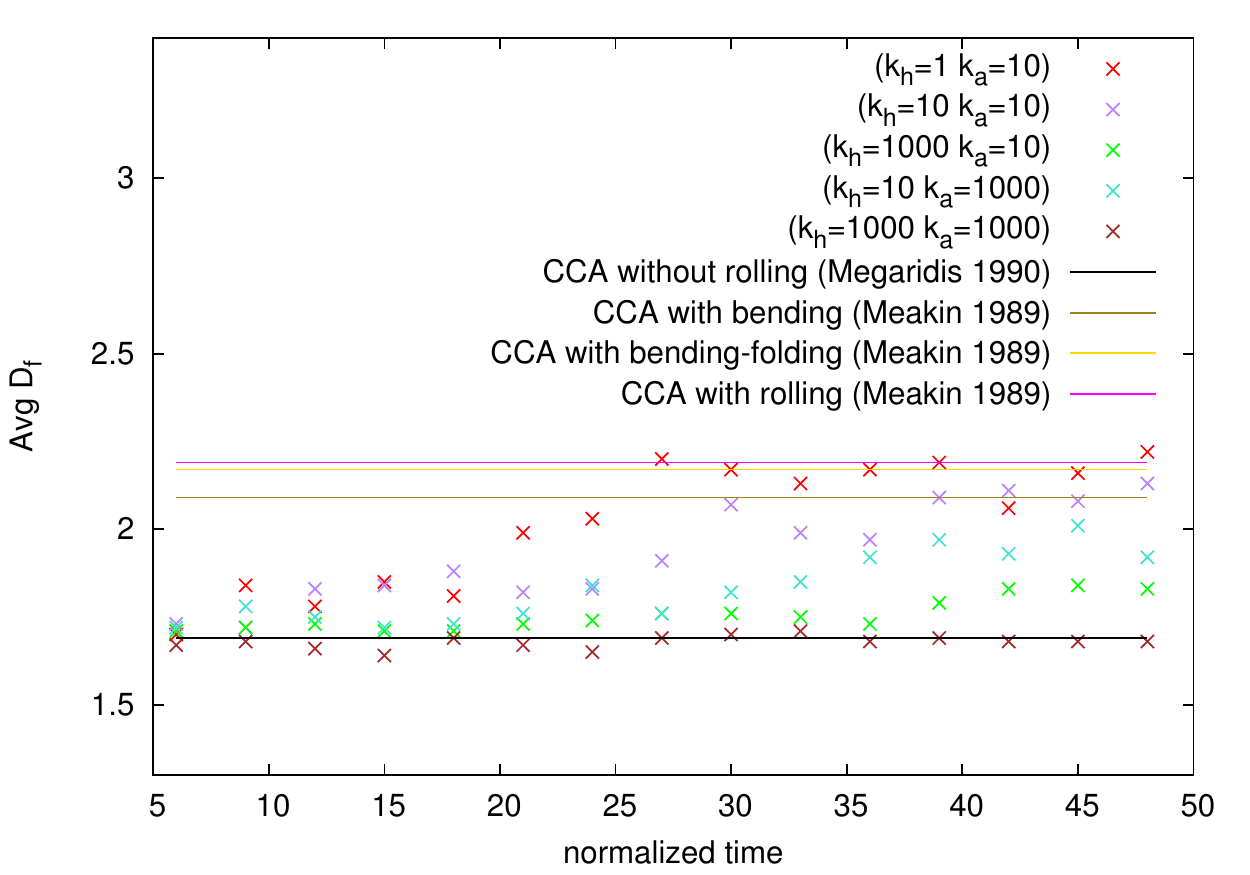}
\label{fig:agg-df}
}
\caption{Dynamics of agglomerates obtained from the AnBV model simulations with varying harmonic ($\tilde{k}_h$) and angular ($\tilde{k}_a$) bond constants. 
\subref{fig:agg-dynamics} Change in the number of agglomerates,
\subref{fig:agg-df} average fractal dimensions of the systems.
}
\label{fig:agglomerates-df}
\end{figure}

The growth and fractal dimension of formed agglomerates ($N>15$) are shown in Figs.~\ref{fig:agg-dynamics} and \ref{fig:agg-df}, respectively, for different $\tilde{k}_h$ and $\tilde{k}_a$ values. At the beginning of the simulations the particles collide and create clusters as indicated by the rapid increase of the number of agglomerates. This rapid increase is followed by a much longer period of reducing the number of agglomerates due to particle-cluster and cluster-cluster collisions that form larger and larger clusters. The dynamics of cluster formation are only little affected by the bond constants chosen for the respective computations. 

The agglomerates' average structures, however, are much more sensitive to the choice of bond constants as illustrated in the same figure and quantified by the different values of $D_f$. The reduced bond constants provide more flexibility, and the agglomerates can bend, fold, elongate and/or shrink during the simulation. The mean fractal dimension varies between $D_f$=1.73 and $D_f$=2.22 for the simulations with smaller bond constants (\ie $\tilde{k}_h \le 10$ and $\tilde{k}_a=10$). This agrees well with the numerical studies by Chen \citeyear{meakin88} and Jullien and Meakin \citeyear{jullien} for cluster-cluster aggregates (CCA) grown by diffusion-limited aggregation (DLA) including impact restructuring. In their studies, the aggregates might fold, bend and twist. The fractal dimension was found to equal $D_f$=2.09 if the agglomerates were allowed to bend, $D_f$=2.17 if both bending and folding were allowed and $D_f$=2.19 if the complete restructuring stages were included for CCA grown by DLA. 

In contrast, the mean fractal dimension varies between $D_f=1.65$ and $D_f=1.71$ for the AnBV model simulation with higher bond constants (\ie $\tilde{k}_h$=1000 and $\tilde{k}_a$=1000). This indicates that no significant restructuring occurs and results match well with the measured fractal dimensions of the cluster-cluster aggregates obtained by various numerical and experimental studies \shortcite{samson,bourrat,zang,megaridis}. It is thus apparent that the correct choice of bond constants allows for the representation of different macroscopic interactions between the particles at the contact point. The effect of the different bond constants can also be quantified in terms of changes in fractal dimension and number of contact points with time as listed for some representative agglomerates in Table~\ref{tab:df-cp}.
%\begin{figure}[H]
%\center{\includegraphics[scale=1.18]{df-cp}}
%\caption{Change in structure of the selected agglomerates ($N_{agg}^1=113$, $N_{agg}^2=130$, $N_{agg}^3=124$ and $N_{agg}^4=136$) during the AnBV model simulations using different bond constants.}
%\label{fig:df-cp}
%\end{figure}
\begin{table}
	\begin{center}
    \caption {Time sequence for changes in $D_f$ and in the number of contact points (cp) of representative agglomerates, A$_i$, during the AnBV model simulations of Case-2. The harmonic and angular bond constants are given as: $A_1$ ($\tilde{k}_h$=1 $\tilde{k}_a$=10), $A_2$ ($\tilde{k}_h$=10 $k_a$=10), $A_3$ ($\tilde{k}_h$=1000 $\tilde{k}_a$=10), $A_4$ ($\tilde{k}_h$=10 $\tilde{k}_a$=1000), $A_5$ ($\tilde{k}_h$=1000 $\tilde{k}_a$=1000). N gives the number of primary particles in each cluster.}
     \begin{tabular}{|l|c|c|c|c|c|c|c|c|c|c|}
    \hline
    $\widetilde{\tau}$ & \multicolumn{2}{c|}{$A_1 (N=136$)} & \multicolumn{2}{c|}{$A_2 (N=130)$} & \multicolumn{2}{c|}{$A_3 (N=140)$} & \multicolumn{2}{c|}{$A_4 (N=124)$} & \multicolumn{2}{c|}{$A_5 (N=136)$} \\ \hline
    ~ & $D_f$ & cp & $D_f$ & cp & $D_f$ & cp & $D_f$ & cp & $D_f$ & cp \\ \hline
    25 & 1.75 & 533 & 2.0 & 524 & 2.0 & 433 & 1.75 & 245 & 1.65 & 280 \\ \hline
    27 & 1.77 & 535 & 2.17 & 527 & 2.0 & 434 & 1.44 & 245 & 1.65 & 280 \\ \hline
    29 & 1.5 & 537 & 2.15 & 528 & 2.0 & 436 & 1.41 & 245 & 1.65 & 280 \\ \hline
    31 & 2.19 & 556 & 2.15 & 528 & 2.0 & 437 & 1.54 & 246 & 1.65 & 280 \\ \hline
    33 & 1.94 & 563 & 2.16 & 528 & 2.0 & 440 & 1.48 & 246 & 1.65 & 280 \\ \hline
    35 & 2.16 & 568 & 2.16 & 529 & 2.0 & 442 & 1.43 & 246 & 1.65 & 280 \\ \hline
    \end{tabular}
    \label{tab:df-cp}
	\end{center}
\end{table}
The computations demonstrate that the resulting morphology depends largely on the harmonic bond potential $\tilde{k}_h$. Small values, \ie $\tilde{k}_h \ll 10$ allow bending of the bond and thus lead to much more compact cluster structures as time advances (cf. agglomerate $A_1$) while large values preserve the gobal structure (cf. agglomerate $A_3$). In addition, high values of $\tilde{k}_a$ support preservation of the agglomerates structure by preventing the particles to fold and thus to create new contact points (cf. agglomerates $A_4$ and $A_5$ with $\tilde{k}_a=1000$ and a nearly constant number of contact points). Note that both $\tilde{k}_a$ and $\tilde{k}_h$ thus allow to control the flexibility and by that the compactness of the agglomerates. However, reducing $\tilde{k}_a$ facilitates a rolling motion which may or may not be physical and will be dependent on the properties of the investigated system/agglomerates. Reducing $\tilde{k}_h$, however, allows the primary particles to separate temporarily, which is unphysical under van der Waals forces. Therefore, reducing $\tilde{k}_a$ should be preferred and $\tilde{k}_h$ should be kept at its physically determined value of $\tilde{k}_h=1000$.

\subsection{Diffusion limited aggregation - Case-3} 

\begin{figure}
\center{\includegraphics[width=.8\textwidth]{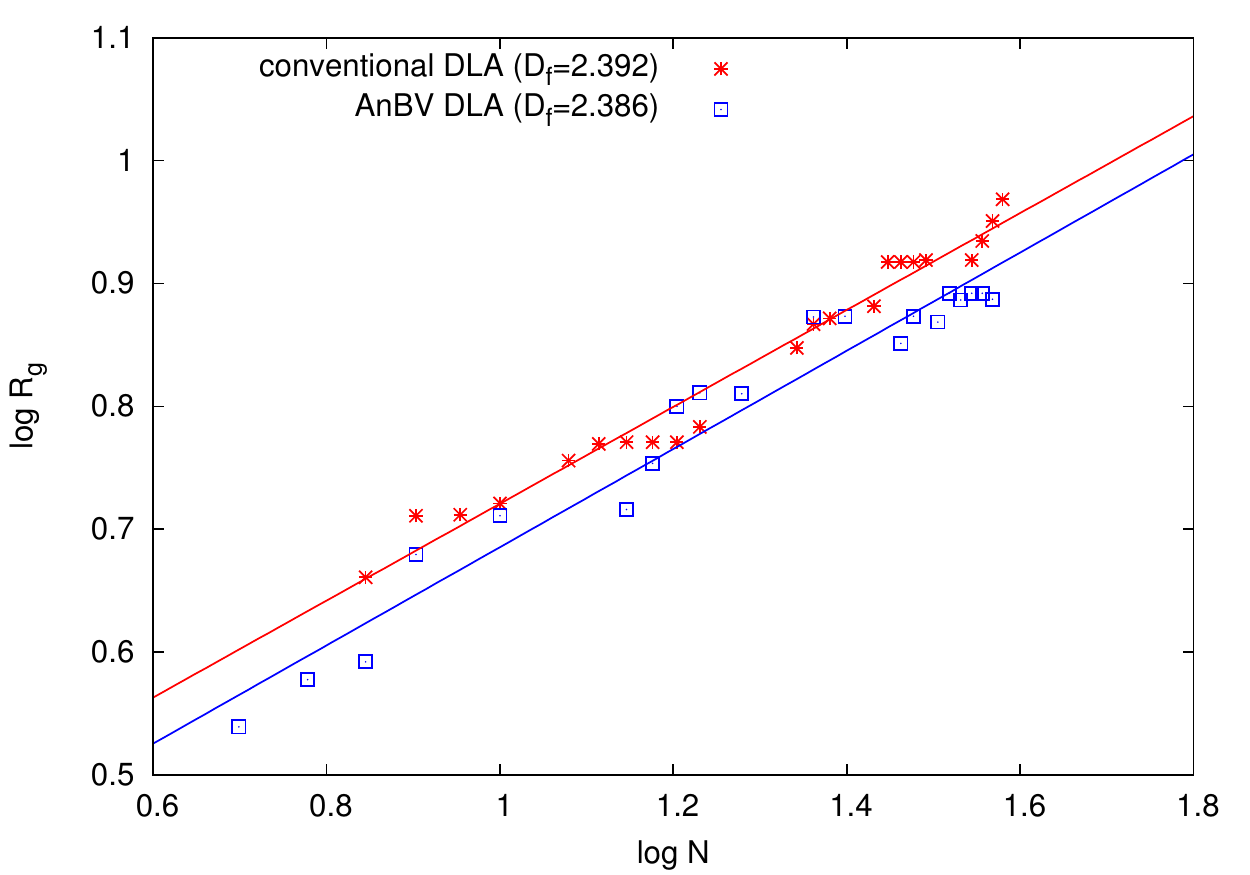}}
\caption{Comparison of the size of the aggregates characterized by radius of gyration, $R_g$ during the growth process. Blues sympols and lines represent the conventional model, pink symols and line represent  the AnBV model. The slopes of the linear fits correspond to the average fractal dimensions of the agglomerates.}
\label{fig:dla}
\end{figure}

The third test case (Case-3) is the simulation of the well-known Diffusion Limited Aggregation (DLA) that can be used to further validate the developed AnBV model. We study DLA of the irreversible growth of a single cluster grown from a seed particle fixed in three-dimensional space. The growth rule is simple, in which the first particle is fixed in the center of the system, and the following particle is then released from a random position far away and is allowed to move due to Brownian motion. If it collides with the first particle, it is connected according to the rules in our AnBV model algorithm and becomes part of the agglomerate. Then, further particles are launched one-by-one and each of them is connected to the cluster after hitting any of the particles belonging to the cluster. 

%Two realisations are compared in Figure~\ref{fig:dla}. 
The growth mechanisms are same for the both cases and explained above. However, in the "conventional DLA" model, when the released particles hit any particle of the cluster, they stop as in the classical DLA model, \ie the formed clusters are immobile. In the "AnBV DLA" model, the aggregate itself is moving, and therefore increasing the probability of collision. The size of the agglomerates are characterized by the radius of gyration,$R_g$ that is given by 
\begin{subequations}
\begin{align}
\label{eq:rg}
R_g&=\sqrt{\frac{\sum_{i=1}^{N}(\textbf{r}_i-\textbf{r}_0)^2}{N}}, \\
\mbox{with}\nonumber \\
\textbf{r}_0&=\frac{1}{N}\sum_{i=1}^{N}\textbf{r}_i,
\end{align}
\end{subequations}
being the agglomerate's center of mass. Figure~\ref{fig:dla} compares $\log(R_g)$ plotted versus $\log(N)$ for the two models. The fractal dimensions of the DLAs can be obtained from linear fits of the data as $D_f=2.386$ and $D_f=2.392$ for the AnBV and the conventional DLA models, respectively. This confirms our expectation that the motion of the cluster has no significant influence on the DLA.

%The fractal dimensions of the DLAs are found by linear fits to be $D_f$=2.386$\pm$0.1 and $D_f$=2.392 for the AnBV and the conventional DLA models. Thus the motion of the cluster has no influence on the DLA, as expected. 
 
\subsection{The growth of soot agglomerates - Case-4}

\begin{figure}
\centering
\includegraphics[scale =0.25] {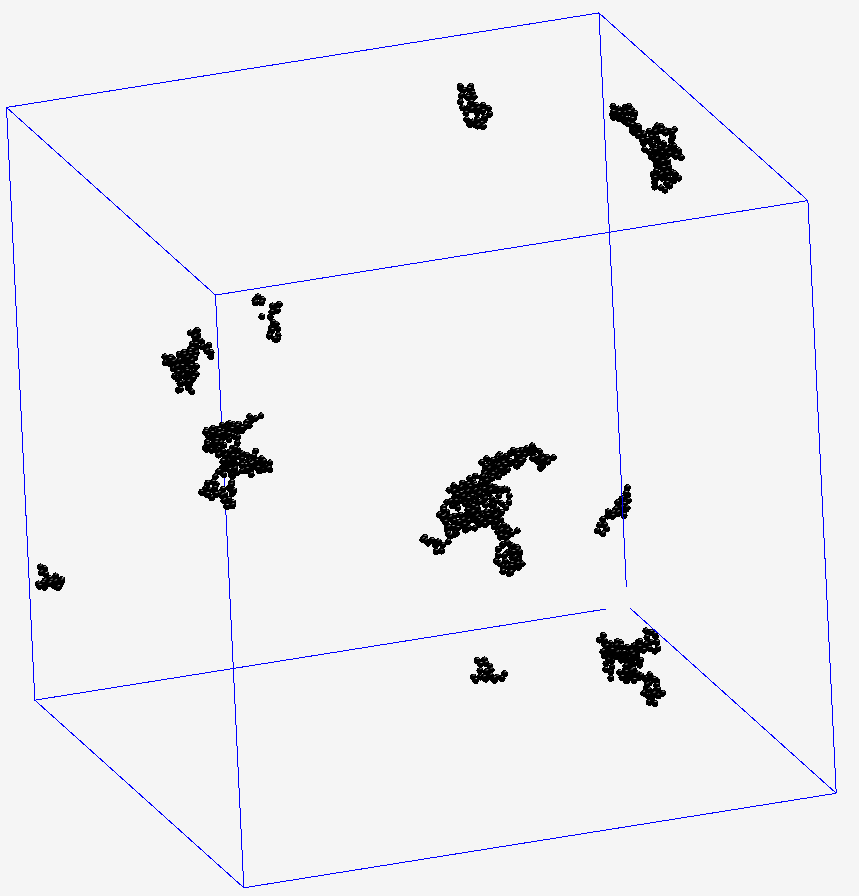}
\label{fig:md-expr-2}
%}
\caption{
Clusters at the end of the LD simulations leading to agglomerates with an average fractal dimension of $D_f$=1.68.}
\label{fig:expr-md}
\end{figure}

For a final validation we compare results from our LD simulations with soot aggregates generated in ethylene diffusion flame and measured by Megaridis \etal \citeyear{megaridis}. They collected samples using a thermophoretic sampling technique by transmission electron microscopy and determined an average fractal dimension of $D_{f,exp}=1.69$. Our LD shall mimic the experimental conditions and we used primary particle diameters of $\sigma=40$ $nm$ and the molecular mean free path length is set to $l_{fluid}=600$ $nm$. The computations show average fractal dimensions of $D_f=1.68$ and analysis is based on all aggregates with more than 15 particles. The maximum agglomerate composes $n=157$ primary particles. The numerical value agrees very well with the value given in Megaridis \etal \citeyear{megaridis} and a qualitative comparison of the shape of the agglomerates (cf. Fig.~\ref{fig:expr-md}) demonstrates the realistic structures of the simulated soot aggregates and supports the need for a novel implementation such as the AnBV model for the simulation of aggregation where particles do not move relative to each other after collision.  
 
\section{Conclusions}
\label{sec:conc}

In this study we studied the capability of four different models to simulate the nanoparticle agglomeration process. All four models require local bonding only. Analyzing the shape and structure of both pre-defined and randomly formed agglomerates in terms of fractal dimension, it became evident that only models with non-central interactions, the so-called AB and AnBV models, are capable of modeling nanoparticle agglomeration without undergoing any restructuring process. Using only non-bonded central interactions (NB model) is not sufficient to hold particles together and to prevent sliding around the contact point of individual particles. Also, forming only harmonic bonds on contact, the SB model, does not fix the particles at their contact points and allows restructuring. Hence, only bulky and compact agglomerates are formed, as one would expect at high process temperatures. We have demonstrated that both the AB and AnBV model can prevent restructuring with tunable degree of compactification. Moreover, the AB and AnBV models can be parameter matched to exhibit identical bending, and thus restructuring properties. However, technically the models are very different. The AB model is computationally faster, but only the AnBV model gives direct access to shear forces, which is required if breakage has to be included. Finally, the AnBV model's ability to reproduce diffusion limited particle-cluster and cluster-cluster agglomerates has been shown. The simulations show remarkable agreement with experimentally determined fractal dimensions of soot aggregates, and we may conclude that the AB and AnBV models provide a suitable implementation of nanoparticle aggregation based entirely on local interactions.

\bibliography{paper}

\end{document}